# Physics-Guided Inverse Design of Optical Waveforms for Nonlinear Electromagnetic Dynamics

Hao Zhang[1,2,*], Jack Hirschman[2,3], Randy Lemons[2], Nicole R. Neveu[2], Joseph Robinson[2], Auralee L. Edelen[2], Tor O. Raubenheimer[2], Dan Wang[4], Ji Qiang[4], Sergio Carbajo[1,2,5,6]

[1]Department of Electrical & Computer Engineering, University of California, Los Angeles, Los Angeles, CA 90095, USA
[2]SLAC National Accelerator Laboratory, Stanford University, Menlo Park, California 94025, USA
[3]Department of Applied Physics, Stanford University, Stanford, CA 94305, USA
[4]Lawrence Berkeley National Laboratory, Berkeley, CA 94720
[5]Physics and Astronomy Department, University of California, Los Angeles, CA 90095, USA
[6]California NanoSystems Institute, Los Angeles, CA 90095, USA
*haozh@g.ucla.edu; scarbajo@ucla.edu

**Abstract:** Structured optical waveforms are emerging as powerful control fields for the next generation of complex photonic and electromagnetic systems, where the temporal structure of light can determine the ultimate performance of scientific instruments. However, identifying optimal optical drive fields in strongly nonlinear regimes remains challenging because the mapping between optical inputs and system response is high-dimensional and typically accessible only through computationally expensive simulations. Here, we present a physics-guided deep learning framework for the inverse design of optical temporal waveforms. By training a light-weighted surrogate model on simulations, the method enables gradient-based synthesis of optical profiles that compensate nonlinear field distortions in driven particle–field systems. As a representative application, we apply the approach to the generation of electron beams used in advanced photon and particle sources. The learned optical waveform actively suppresses extrinsic emittance growth by more than 52% compared with conventional Gaussian operation and by approximately 9% relative to the theoretical flattop limit in simulation. We further demonstrate experimental feasibility by synthesizing the predicted waveform using a programmable pulse-shaping platform; incorporating the measured optical profile into beamline simulations yields a 31% reduction in the extrinsic emittance contribution. Beyond accelerator applications, this work establishes a general way for physics-guided inverse design of optical control fields, enabling structured light to approach fundamental performance limits in nonlinear photonic and high-frequency electromagnetic systems.

1. Introduction

The ability to engineer the temporal structure of light has become a central capability of modern photonics[1,2]. Structured optical waveforms provide a versatile means to control nonlinear light–matter interactions and drive complex dynamical systems, enabling applications ranging from coherent quantum control and attosecond science to laser materials processing and advanced radiation sources. Structured optical waveform synthesis has long been a cornerstone of modern photonics, enabling precise control over light-matter interactions ranging from coherent quantum control to ultrafast laser processing[2–7]. For example, in the realm of high-brightness electron sources, the laser pulse plays an even more fundamental role: it acts as the initial template that defines the birth of the electron bunch[7–9]. For next-generation scientific instruments such as X-ray Free-Electron Lasers (XFELs) and single-shot Ultrafast Electron Diffraction (UED) systems, the quality of this electron generation directly dictates the ultimate brightness and coherence of the final probe beam[3,10–12]. This quality is fundamentally governed by the transverse emittance of the electron beams, which measures the spread of electron positions and momenta in phase space. A lower emittance implies a more laminar and focusable beam, which is essential for achieving Angstrom-level spatial resolution in UED and maximizing the peak brilliance in XFELs. Ideally, the transverse emittance of the electron beam should be bounded only by the intrinsic thermal emittance of the photocathode, which is a fundamental thermodynamic limit[13–16].

However, approaching this thermal limit in high-brightness regimes is fundamentally constrained by nonlinear extrinsic field distortions[17,18]. As electrons are emitted and accelerated, they experience time-dependent and non-uniform transverse forces, arising from phase-dependent RF fields and non-uniform transverse forces act as a dynamic lens, actively distorting the beam phase-space and degrading emittance far beyond the intrinsic thermal limit, irreversibly degrading the beam quality. Conventionally, this issue is addressed by structured optical waveform synthesis into a flattop temporal profile at the photocathode[19,20]. Theoretically, laser temporal shaping aims to linearize the effective field forces experienced by the bunch, thereby minimizing the extrinsic emittance growth during the initial extraction[21–23]. By generating a more uniform longitudinal density, the flattop profile seeks to align the phase-space ellipses of different bunch slices. This intuitive approach is based on simplified models that neglect the complex dynamic evolution of the electron bunch. As the beam accelerates, relativistic effects, non-ideal field curvatures, and time-dependent focusing forces cause the phase-space distribution to reconfigure and deviate significantly from the initial state, rendering the static flattop solution suboptimal for minimizing the final extrinsic emittance. The search for the optimal temporal shape thus becomes an inverse problem in a high-dimensional parameter space that is heavily dependent on the specific machine configuration, rendering the process computationally impractical for traditional particle-in-cell (PIC) simulations[24–26].

In recent years, machine learning (ML) and deep learning (DL) have emerged as a transformative approach for modeling complex nonlinear physical systems, offering orders-of-magnitude computing time acceleration over traditional numerical solvers[27–32]. These models are typically trained via supervised learning on extensive datasets generated by high-fidelity simulations, effectively capturing the high-dimensional mapping between control parameters and beam phase-space metrics[26,32–34]. To address the issue of machine-configuration dependence, these approaches can incorporate experimental measurements (often via transfer learning) to calibrate the model against real-world machine states, although this is typically constrained by the scarcity of experimental training data[27,28,35,36]. By approximating intensive simulations with pre-trained models, neural networks enable the exploration of vast parameter spaces that were previously inaccessible. Beyond simple prediction, the inherent differentiability of these neural architectures has enabled a new paradigm of inverse design, where optimal system inputs can be directly synthesized by backpropagating gradients through the model[37,38]. However, pure data-driven approaches can sometimes yield unphysical solutions. To address this, physics-informed deep learning strategies have been developed to embed fundamental physical laws, such as conservation principles and boundary conditions, directly into the network architecture or loss functions, ensuring that the optimized results remain physically realizable[39–41].

To overcome the computational bottleneck of classical methods[42–44], we propose a general-purpose, physics-informed deep learning approach for the inverse design of temporal optical waveforms. Recent advances in programmable pulse shaping, nonlinear frequency conversion, and structured-light synthesis now enable optical waveforms with unprecedented temporal complexity. Unlike traditional forward optimization methods that rely on trial-and-error searches within limited parameter sets (e.g., stacking Gaussian pulses), our approach adopts a high-fidelity differentiable surrogate model to establish a bidirectional map between the laser waveform and the electron beam dynamics[37]. This formulation allows us to treat the temporal intensity profile not as a rigid function, but as a flexible, continuous variable that can be tuned to counteract complex field-induced distortions. Conceptually, we view this drive optical waveform as a "photonic fingerprint" imprinted onto the electron bunch[9,45,46]. While standard Gaussian or flattop profiles represent generic initial conditions that allow nonlinear distortions to accumulate, our approach identifies unique, high-resolution characteristics, specifically subtle modulations at the pulse edges, that effectively mitigate the specific signature of extrinsic field aberrations at the source.

In this paper, we demonstrate that this optimized waveform acts as a precise key to unlock the intrinsic thermal limit of the electron source. By evaluating the system in a high-brightness regime, our method isolates and suppresses the extrinsic emittance growth by over 52% compared to standard operations, and by ~9% compared to the theoretically preferred flattop baseline. We also experimentally generated a similar optical waveform, which resulted in a robust 31% reduction in the extrinsic emittance contribution compared to the Gaussian-profile-based standard mode in simulation. This work demonstrates how inverse-designed optical drive fields can approach fundamental limits in complex nonlinear photonic systems, with high-brightness electron sources serving as a representative platform.

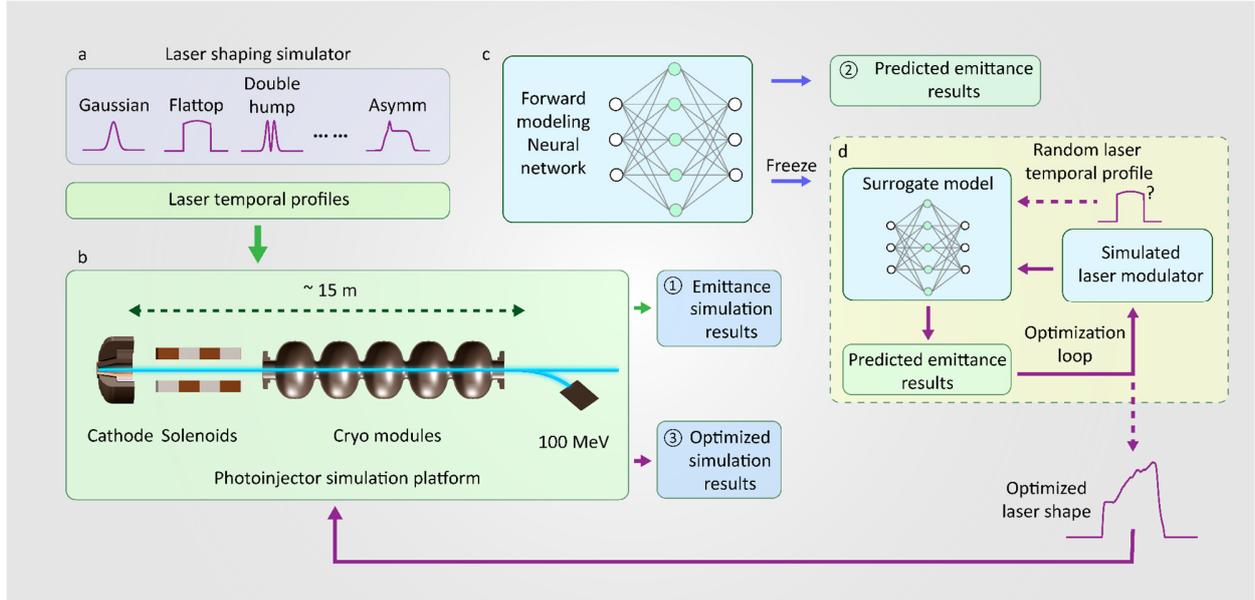

**Figure 1. Physics-informed inverse ultrafast laser temporal shaping workflow for minimizing extrinsic emittance contribution.** Throughout the workflow, the beamline configuration remains fixed, with the laser temporal profile serving as the sole variable. (a) A diverse set of candidate ultrafast laser temporal profiles: Gaussian, flattop, double-hump, asymmetric, etc., is generated by the laser-shaping simulator. (b) Each profile drives the photoinjector simulation chain (cathode → solenoids → cryogenic accelerating modules, ~15 m total, up to 100 MeV) to produce photoinjector emittance results: initial emittance mappings ①. (c) A forward-model neural network is trained on these simulation pairs to predict beam emittance from an arbitrary optical waveform; once trained, its weights are frozen to serve as a fast surrogate model ②. (d) In the inverse-learning loop, the frozen surrogate is paired with a differentiable laser-modulator emulator. Starting from a random initial profile, gradients backpropagate through the surrogate to iteratively update the laser shape, yielding the optimized temporal profile, which is then re-evaluated in the full photoinjector simulator ③.

## 2. Results and discussion

**Approach overview.** The overall architecture of the proposed physics-informed inverse design approach is schematically illustrated in Figure 1. The primary objective is to identify the optimal temporal optical waveform $I(t)$ that minimizes the transverse emittance of the electron beam at the injector exit. Unlike conventional optimization approaches that parameterize the laser pulse with a few scalar descriptors (e.g., FWHM or skewness), we treat the discretized intensity profile as a high-dimensional vector $x \in \Re^N$ (where $N = 301$). By viewing the intensity at each time step as an independent degree of freedom, we expand the optimization landscape into a high-dimensional parameter space.

Our approach proceeds in three interconnected stages. First, a high-fidelity data generation phase (Figure 1a-b) constructs a comprehensive knowledge base. A well-designed library of temporal profiles, ranging from standard operational mode (Gaussian-based) to complex structured waveforms, is fed into the Lightsource unified modeling environment (LUME)-Impact-T simulation platform[47]. This establishes a ground-truth mapping between the input laser shapes and their corresponding beam dynamics, capturing the intricate physics of extrinsic-field-influenced transport and the resulting phase-space distortions.

Second, this dataset is used to train a deep neural network that serves as a differentiable surrogate model (Figure 1c and Figure 2a). By learning the non-linear mapping from the high-dimensional optical waveform to the scalar emittance value, the neural network effectively acts as a "virtual optical-field-driven beam dynamics," replacing the computationally expensive PIC solvers (~7 minutes with a 32-core CPU, tracking 300k particles on a 32 × 32 × 32 grid) with a millisecond-scale inference engine.

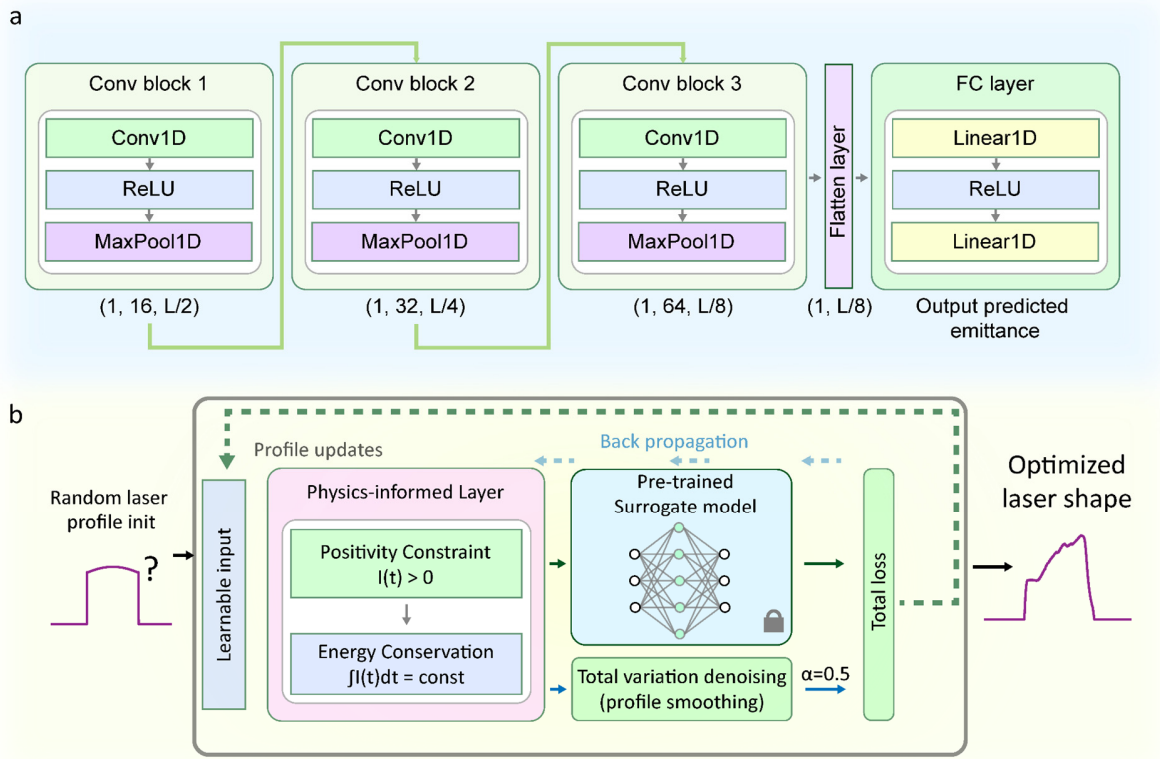

**Figure 2. Supervised neural network architecture and physics-informed inverse shaping loop.** (a) Forward-modeling network architecture used as a surrogate to predict beam emittance from a discretized laser temporal profile of length L. Three sequential 1D convolutional blocks (Conv1D → ReLU → MaxPool1D) expand channel depth from 1 → 16 → 32 → 64 while downsampling the time axis by a factor of 2 each. The pooled output is flattened and passed through two fully connected (Linear1D → ReLU → Linear1D) layers to yield the predicted emittance. (b) Physics-informed inverse-learning approach. A random initial optical waveform is treated as a learnable tensor and first passed through a customized "physics-informed layer" that enforces positivity ($I(t) > 0$) and energy conservation ($\int I\, dt$ constant). The processed profile is then evaluated by the frozen, pre-trained surrogate to compute an emittance loss. A total-variation (TV) regularization term, defined as the L1-norm of the profile's temporal gradient, is added to the loss function. The weighting hyperparameter $\alpha = 0.5$ specifically governs the micro-smoothness of the profile; it was empirically tuned to suppress unphysical high-frequency fluctuations (noise) while preserving the necessary macroscopic features for emittance minimization. Gradients of the combined loss are back-propagated through the frozen surrogate model to update the optical waveform iteratively, yielding the optimized temporal shape.

The inverse design is executed via a gradient-based optimization loop (Figure 1d and Figure 2b). We fix the weights of the pre-trained surrogate to establish a stable, differentiable landscape and couple it with a physics-constraint layer that mathematically enforces hardware limitations (positivity and energy conservation). Starting from a random initialization, the optimal profile is iteratively sculpted by backpropagating the gradients of the emittance loss function through the frozen surrogate model to the input layer. This allows the optimizer to efficiently traverse the high-dimensional parameter space and converge to a highly optimized solution, which is subsequently benchmarked against the full physics simulation to confirm the predicted emittance reduction.

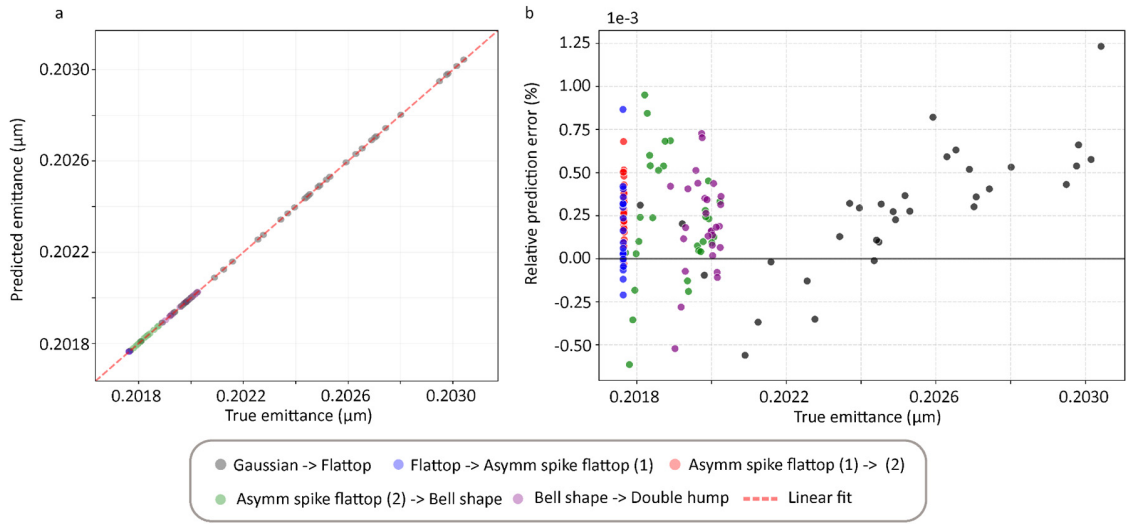

**Figure 3. Surrogate model emittance prediction accuracy along continuous laser-shape morphing paths.** Simulations were performed with the laser temporal profile acting as the sole variable. (a) Scatter of predicted vs. reference simulation results for profile samples taken along five continuous transitions in laser temporal shape: Gaussian→Flattop (black), Flattop→Asymm-Spike-Flattop (v1) (blue), Asymm-Spike-Flattop (v1)→Asymm-Spike-Flattop (v2) (red), Asymm-Spike-Flattop (v2)→Bell (green), and Bell→Double-Hump (purple). The dashed red line is the ideal y=x fit. (b) Absolute prediction error plotted against true emittance for the same samples, with each color denoting one morphing path and the horizontal dashed line at zero error.

Prior to inverse design, we validated the surrogate model's ability to generalize beyond the training anchors by evaluating its accuracy along continuous morphing paths between distinct pulse topologies. Figure 3 presents the comparison between the reference simulation results (LUME-Impact-T) and the surrogate predictions for five transition trajectories. The scatter plot in Fig. 3(a) demonstrates a robust linear agreement between predicted and simulated values, with a regression slope of ~0.99. This indicates that the model correctly reproduces the physical scaling laws across the dynamic range without significant bias.

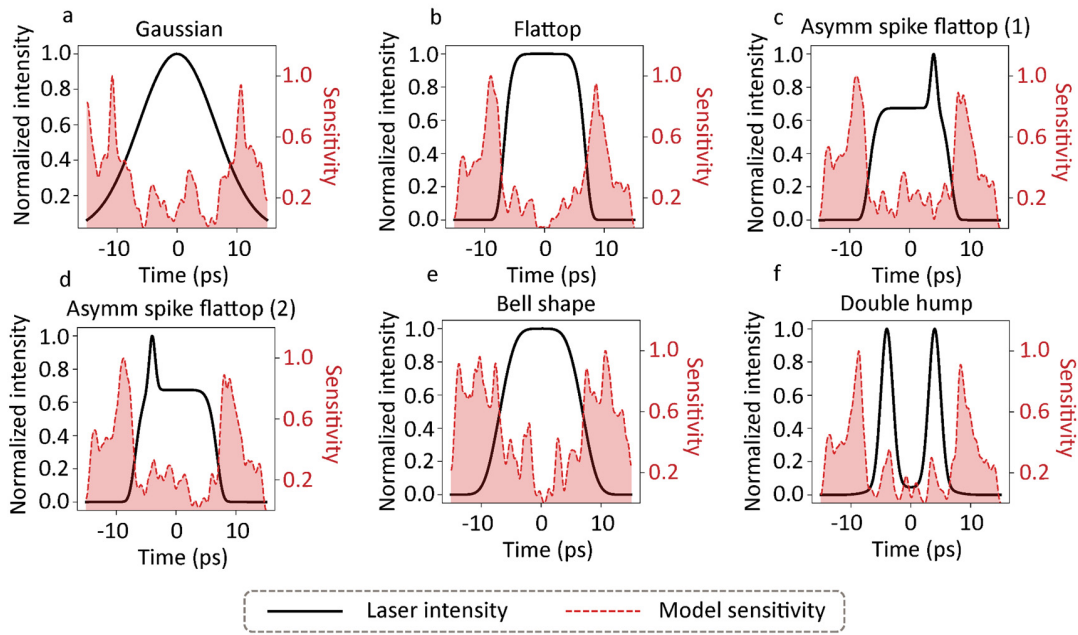

**Figure 4. Sensitivity maps for learning representative laser temporal profiles.** Panels (a–f) show six pulse shapes: Gaussian, Flattop, Asymm Spike Flattop (1), Asymm Spike Flattop (2), Bell Shape, and Double Hump, which are plotted as solid black curves (normalized intensity vs. time). Overlaid in each panel is the model's saliency for that profile, defined as the absolute gradient of the frozen surrogate's emittance prediction with respect to the input intensity, smoothed by a Gaussian filter (σ = 3) and normalized to [0,1]. The red shaded area and dashed red outline represent the temporal locations where small perturbations of the pulse shape have the greatest impact on predicted emittance. Time is centered at the pulse's mean arrival ⟨t⟩ and spans ±15 ps.

The transition from Gaussian to Flattop profiles (black markers) exhibits the largest variation in emittance, corresponding to the shift from a regime of significant extrinsic phase-space distortion to a regime of enhanced temporal matching. The surrogate successfully tracks this broad evolution. Furthermore, for complex structured profiles, such as the Asymm-Spike-Flattop (red) and Bell shapes (green), the model retains sufficient sensitivity to resolve emittance shifts caused by minor topological adjustments. While Fig. 3(b) shows minor residuals, the consistently high correlation confirms that the CNN extracts the critical temporal features governing field-induced beam quality, establishing its reliability for the optimization stage.

To gain insight into the feature selection process of the neural network, we performed a sensitivity analysis[48] using gradient-based sensitivity maps as described in the method section. Figure 4a-f visualizes the gradient of the predicted emittance with respect to the input intensity overlaid on six representative pulse profiles. The red-shaded regions indicate the temporal domains where intensity perturbations are identified by the model as having a measurable impact on the projected emittance.

For the Gaussian profile (Figure 4a), the sensitivity map reveals a broad distribution. While significant gradients are observed at the pulse wings (determining the effective bunch length and its overlap with RF curvature), a noticeable sensitivity region persists near the pulse center. This suggests that the model accounts for the temporal phase alignment, which directly influences the integrated interaction with the accelerating fields. In contrast, for the Flattop (Figure 4b), the sensitivity is more distinctly confined to the rising and falling edges, confirming that for plateau-like distributions, the boundary steepness becomes the dominant optimization variable for linearizing the phase-space response over the flattop amplitude.

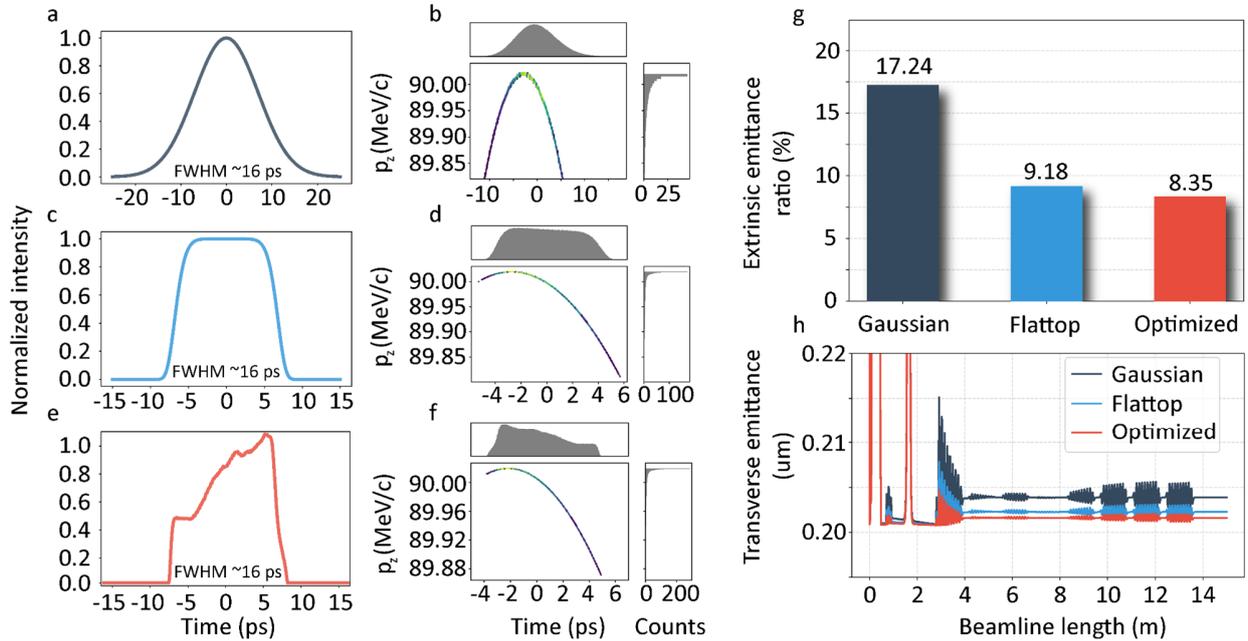

**Figure 5. Comparative beam dynamics for Gaussian, Flattop, and Optimized laser temporal profiles calculated by the IMPACT simulations.** (a,c,e) Show the normalized laser intensity vs. time for (a) a standard Gaussian pulse, (c) an ideal flattop pulse, and (e) the physics-informed optimized pulse. In (b,d,f), each corresponding profile's beamline simulation results are displayed as a joint distribution of final longitudinal momentum $p_z$ vs. relative arrival time $t-\langle t \rangle$ (center), with the top histogram giving the temporal arrival distribution and the right histogram showing the marginal $p_z$ distribution. (g) Compares the resulting extrinsic emittance contribution and (h) transverse emittance dynamic along the beamline (max: 15 m, ~100 MeV) for the three cases: Gaussian, ideal flattop, and optimized. Note that the accelerator lattice parameters (e.g., solenoid fields and RF settings) were fixed to values pre-optimized for the Gaussian baseline to isolate the performance gain attributable solely to temporal shaping.

In the case of structured profiles, such as the Asymmetric Spike variations (Figure 4c, d), the model demonstrates a topology-aware response. The high-saliency regions are observed to coincide with two specific features: the steep rising/falling transitions and local intensity protuberances (sudden bulges). The strong response to the edges indicates the model's focus on the longitudinal slice matching, while the sensitivity to the local convexities implies that the surrogate is capturing the impact of localized temporal modulations. These local fluctuations may act as sources of nonlinear phase-space distortions, and the model correctly identifies them as critical areas requiring refinement during the inverse design.

Figure 5 compares the beam dynamics performance of three distinct laser temporal profiles: a standard Gaussian reference (Figure 5a), an ideal Flattop (Figure 5c), and the physics-informed AI-optimized profile (Figure 5e). The corresponding longitudinal phase-space distributions at the injector exit are visualized in Figure 5b, d, and f, respectively.

The standard Gaussian pulse (Figure 5a, RMS Length 2.1 mm) serves as the first (practical) baseline. As shown in Figure 5b, the resulting phase-space exhibits a noticeable nonlinear curvature, stemming from the phase-dependent field effects sampled by the non-uniform Gaussian temporal distribution. This nonlinearity precludes effective compensation via linear optics, resulting in a residual extrinsic emittance growth of 17.24% relative to the intrinsic thermal emittance for an electron bunch with an rms length of 1.26 mm.

As expected, transitioning to the Flattop profile, our second (theoretical) baseline (Figure 5c, RMS length 1.4 mm), reduces the overall emittance growth. By providing a uniform temporal distribution, the Flattop pulse homogenizes the phase-dependent focusing forces within the bunch core, aligning with theoretical predictions for optimal phase-space matching[21,22]. The phase-space distribution in Figure 5d exhibits a linearized structure compared to the baseline, suppressing the residual extrinsic emittance growth to 9.18% relative to the intrinsic thermal emittance for an electron bunch with an rms length of 0.87 mm.

The optimized profile generated by the proposed method (Figure 5e, RMS length 1.4 mm) further improves this performance. Beyond the linearity achieved by the flattop pulse, the optimized shape incorporates specific modulation slopes that actively compensate for residual higher-order nonlinearities. The impact is clearly observed in the marginal momentum distribution in Fig. 5(f), where the spectral density in the high-energy region exhibits a sharp, pronounced peak, reaching nearly 240 counts, which is ~33% greater than the ideal flattop case and ~220% compared with the standard mode. This peak signifies a high degree of phase-space linearization, where a larger fraction of electrons is compressed into a narrow energy band. Figure 5(g,h) summarizes the emittance evolution, which shows the optimized profile reduces the residual extrinsic emittance growth ratio to 8.35% µm relative to the intrinsic thermal emittance for a 0.86 mm (rms) electron bunch. This represents a substantial reduction of ~52% compared to the Gaussian baseline and a further ~9% improvement over the Flattop benchmark. This improvement is likely attributable to the asymmetric profile, which is tailored to compensate for the time-dependent field curvature encountered during the initial acceleration phase. By redistributing higher photon density towards the tail, the optimized profile counteracts the phase-dependent focusing variations that typically distort the transverse phase-space, facilitating the formation of a more linear longitudinal-to-transverse mapping. This improvement is promising because the initial source quality sets the fundamental brightness limit for the entire accelerator complex. Minimizing extrinsic source emittance is a cross-cutting priority essential for advancing the missions of basic energy sciences, high energy physics, and fusion energy sciences, directly enabling higher brightness and spatiotemporal resolution for future scientific discoveries[46,49].

**Experimental Feasibility Analysis.**

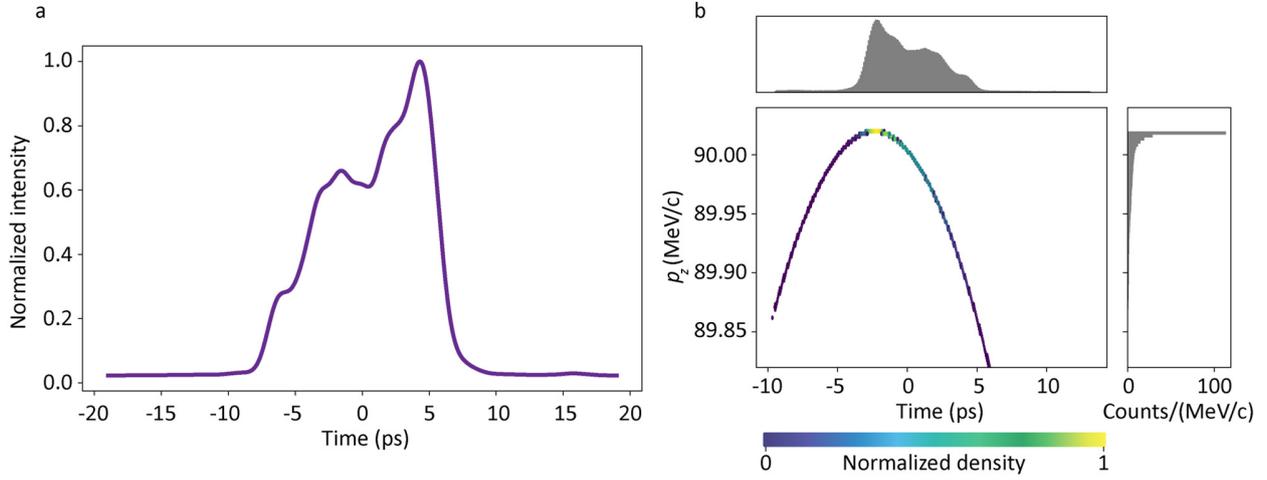

**Figure 6. Experimental feasibility validation.** (a) The experimentally synthesized pulse (RMS Length: 3.72 ps) was generated via the dispersion-controlled nonlinear synthesis (DCNS) method. The measured profile approximates the feature of the optimized design shown in Figure 5e. (b) Corresponding beam dynamics simulation result using the measured experimental optical waveform.

Translating the theoretically optimized pattern into physical reality involves overcoming two primary optical challenges, for which established solutions have been recently demonstrated[50,51].

Standard programmable pulse shapers are typically designed for broadband femtosecond pulses and lack the spectral resolution required to sculpt the narrowband picosecond pulses (>10 ps) used in photoinjectors. Meanwhile, direct programmable shaping in the UV regime is further hindered by the scarcity of efficient, damage-resistant modulator materials[52]. This bottleneck can be effectively circumvented through dispersion-controlled nonlinear synthesis (DCNS) method[6,22] or indirect shaping strategies, such as the previously reported four-wave mixing scheme in gas-filled hollow-core waveguides[53–55]. By transferring complex modulation patterns from the infrared (IR) to the UV via nonlinear mixing processes, arbitrary UV waveforms can be generated without subjecting delicate optics to high-energy photons.

To validate the practical viability of our inverse design approach, we utilized a DCNS-based setup to experimentally synthesize an optical waveform that approximates the key features of the AI-optimized solution, as shown in Figure 6a (RMS length 1.12 mm). We then imported this measured UV profile into the simulation to assess the realistic beam quality against a nominal operational baseline (a standard Gaussian distribution). Despite the deviation from the idealized theoretical curve, the experimentally shaped pulse yields an extrinsic emittance reduction of ~31% and ~47% improvement in peak projected energy density for the generated electron bunch (rms length: 0.88 mm) relative to the baseline. Furthermore, the longitudinal phase-space analysis (comparing with Figure 5a,b) confirms that the optimized shaping successfully concentrates the electron bunch density. As evidenced in the marginal distribution (Figure 6b), the spectral density peak exceeds 110 counts and is localized in the high-energy region.

Given the demonstrated efficacy in simulation, the lightweight and millisecond-scale inference speed of the surrogate model provides the necessary prerequisites for extending this strategy to broader experimental and design scenarios. We identify two promising avenues: real-time digital twins for online feedback, and source-design optimization for next-generation high-brightness facilities. Unlike computationally intensive PIC simulations, the proposed neural network enables rapid inference suitable for online control systems. The inverse model could be integrated into the laser control loop to dynamically adjust programmable pulse shapers in response to shot-to-shot jitter in bunch charge or RF phase. This capability would serve as an active stabilizer, maintaining consistent beam quality against environmental drifts without interrupting operation[7,56,57].

Furthermore, this method holds significant potential for compact Inverse Compton Scattering (ICS) light sources. To maximize photon flux, these sources must operate under conditions where extrinsic emittance growth typically incurs a penalty in spectral bandwidth. Our results suggest that optimized temporal profiles can compensate for complex phase-space distortions, enabling high-flux X-ray generation with narrow spectral bandwidths for medical and industrial applications[9,46,58]. Similarly, for continuous-wave (CW) XFELs targeting the intrinsic thermal emittance limit, this source-side optimization effectively minimizes emittance growth[59–61]. Implementing such strategies could directly enhance X-ray peak brightness and potentially relax undulator design requirements, improving the overall performance efficiency of future photon science facilities.

In summary, we have demonstrated a complete, lightweight physics-informed inverse design method that overcomes the computational barriers of high-dimensional structured optical waveform synthesis. By using a pre-trained convolutional surrogate model, we successfully identified non-intuitive temporal profiles that reduce transverse emittance by 76% when compared with the standard operation mode, surpassing the theoretical performance of standard flattop pulses in the nominal beamline configuration by ~17%, all while maintaining the accelerator lattice fixed at its Gaussian-optimized settings. Our analysis reveals that the optimized optical structure actively mitigates non-ideal field-gradient effects and the resulting phase-space broadening at the cathode.

While the remarkable computational efficiency of this framework readily enables the real-time, autonomous stabilization strategies discussed above, its implications extend far beyond electron beam dynamics. This work illustrates how programmable light fields, combined with physics-guided machine learning, can transform structured optical waveform synthesis into a general-purpose control strategy for nonlinear photonic and electromagnetic systems. Such approaches will enable the systematic discovery of optimal optical control fields for a wide array of applications, ranging from ultrafast spectroscopy and plasma accelerators to advanced radiation sources.

3. **Methods**

**Optical waveform generation.** We generated a diverse library of temporal optical waveforms ranging from standard Gaussian to theoretical flattop distributions, as well as complex structured waveforms. The temporal domain was defined over a window of $t \subseteq [-15,15]$ ps, discretized into $N_t = 301$ grid points. We established a characteristic time scale $\sigma_t$ corresponding to a full width at half maximum (FWHM) of 15 ps for a standard Gaussian pulse, defined as $\sigma_t = FWHM/(2\sqrt{2ln2})$.

A set of six foundational profiles, denoted as $S_k(t)$, was defined to span the temporal space of interest:

**Standard (Gaussian):**
$$S_G(t) = exp[-t^2/(2\sigma_t^2)]$$

**Flattop:** Approximated by an 8th-order Gaussian:
$$S_{flattop}(t) = exp[-t^8/(2\sigma_t^8)],$$

**Bell-shaped:** An intermediate 4th-order Gaussian:
$$S_{Bell}(t) = exp[-t^4/(2\sigma_t^4)].$$

**Asymmetric Structures:** To mimic realistic laser imperfections or explore non-trivial compensation mechanisms, we introduced profiles with localized perturbations. These included a super-Gaussian with a trailing spike ($t = +4\ ps$) and a leading spike ($t = -4\ ps$), modeled by superposing a narrow Gaussian perturbation ($\sigma_p = 0.5\ ps$) onto the flattop base.

**Double-hump:** A bimodal distribution constructed from two displaced Gaussians.

To expand these base shapes into a continuous dataset, we employed a homotopic interpolation strategy. A total of 1000 unique temporal profiles were generated by linearly morphing between consecutive anchor shapes. For a transition between shapes $S_i$ and $S_{i+1}$ (an intermediate profile $S(t, \alpha)$) is calculated as:

$$S(t, \alpha) = (1 - \alpha)S_i(t) + \alpha S_{i+1}(t)$$

$\alpha \subseteq [0,1]$ is the interpolation factor. All generated profiles were normalized to ensure the independence of pulse shape from total pulse energy (or bunch charge), which is handled as a separate scaling parameter in the simulation.

**Photoinjector simulation platform.**

Numerical experiments were conducted using the LUME-Impact-T framework[47,62], which solves the 3D Poisson equation to accurately capture the evolution of beam dynamics under the influence of complex, time-dependent electromagnetic fields and collective effects. The simulation tracks $3\times10^5$ particles generated via a Hammersley sequence. We consider a standard high-brightness injector layout comprising a 185.7 MHz VHF gun, a 1.3 GHz buncher, and a 1.3 GHz superconducting booster linac.

To isolate the impact of temporal pulse shaping, the transverse optical waveform was fixed as a uniform distribution with a radius of R=0.5 mm. To evaluate the performance gain attributable solely to temporal shaping under a nominal operating condition, the accelerator lattice parameters (e.g., solenoid fields and RF settings) were fixed to values pre-optimized for a standard Gaussian laser baseline. The photoemission process was modeled as instantaneous, as the characteristic response time of the Cs2Te photocathode (< 1 ps) is negligible compared to the pulse duration. The initial thermal spread was modeled with a Mean Transverse Energy (MTE) of 330 meV. These parameters define a baseline thermal emittance of ~0.20 μm. The mesh resolution for the solver was set to 32×32×32. Table I lists the simulation configurations used for data generation and validation.

Table I. Simulation configuration.

| Component | Parameter | Value | Unit |
|---|---|---|---|
| VHF Gun | Frequency | 185.7 | MHz |
| | Peak On-axis Field (Ez) | 20.0 | MV/m |
| | SOL1 | 0.056 | T |
| L-Band Buncher | Frequency | 1.3 | GHz |
| | Peak Field Gradient | 1.8 | MV/m |

| | Frequency | 1.3 | GHz |
|---|---|---|---|
| **Superconducting Linac** | Capture Gradient (Cavity 1) | 12.5 | MV/m |
| | Velocity Bunching Gradient (Cavity 2) | 3.1 | MV/m |
| | Main Acceleration Gradient (Avg, Cavity 5-8) | 16.0 | MV/m |

**Emittance calculation**

The total normalized transverse emittance of the electron beam, $\varepsilon_{total}$, is a critical figure of merit for beam quality. In our analysis, we treat the total emittance as the quadrature sum of independent contributions. We decouple the initial thermal emittance at the cathode from the extrinsic emittance growth induced by non-ideal field effects during beam transport.

Under the assumption that the thermal emission process and the subsequent extrinsic field-driven dynamics are uncorrelated mechanisms, the total normalized emittance[63] can be simplified as:

$$\varepsilon_{total} = \sqrt{\varepsilon_{int}^2 + \varepsilon_{ex}^2}$$

where $\varepsilon_{int}$ represents the intrinsic emittance and $\varepsilon_{ex}$ denotes the extrinsic emittance. The intrinsic emittance is determined by the emission physics at the photocathode surface. It sets the fundamental lower limit for the beam emittance and is defined by the laser spot size and the mean transverse energy (MTE) of the emitted electrons. It is calculated using the following expression:

$$\varepsilon_{int} = \sigma_x \sqrt{\frac{MTE}{m_e c^2}}$$

where $\sigma_x$ is the root-mean-square transverse size of the laser spot on the cathode. $MTE$ is the Mean Transverse Energy of the electrons upon emission. $m_e c^2$ is the electron rest mass energy (~511 KeV).

**Forward modeling neural network.**

The implemented 1D Convolutional Neural Network (CNN) as the surrogate model is shown in Figure. 2a. The network takes a discretized intensity vector $x \in \Re^{301}$ as input and predicts the scalar projected emittance. We selected a convolutional architecture to exploit the inherent locality and translation invariance of the pulse shaping physics; critical features such as sharp rise times and local modulations are best captured by sliding kernels rather than dense connections. The architecture proceeds in two stages:

**Stage-1. Hierarchical Feature Extraction:** Three sequential convolutional blocks (Conv1D → ReLU → MaxPool1D) progressively expand the channel depth from 1 to 16, 32, and finally 64. This design extracts increasingly complex geometric features while downsampling the temporal dimension by a factor of 2 at each stage to reduce computational redundancy.

**Stage-2. Regression:** The resulting high-level feature map is flattened into a latent vector and passed through a regression head comprising two fully connected layers.

**Physics-guided inverse learning.**

Once the surrogate model is well trained, its weights are frozen and embedded into the inverse optimization loop as illustrated in the Figure. 2b. In this configuration, the discrete optical waveform $x$ is initialized as a flattop profile and treated as the learnable parameter. To ensure that the mathematically optimized solutions correspond to physically realizable laser pulses, the raw input tensor is first processed through a customized physics-informed layer prior to entering the surrogate. This layer strictly enforces a non-negativity constraint ($I(t) \geq 0$) to prevent unphysical negative light intensities and simultaneously normalizes the profile to maintain a constant integrated pulse energy ($\int I(t) = Const$), thereby decoupling the temporal shape optimization from total bunch charge variations.

The optimization is driven by a composite loss function $L_{total} = L_{emittance} + L_{TVD}$, where $L_{emittance}$ is the beam emittance predicted by the frozen surrogate model. To mitigate high-frequency numerical noise and ensure spectral smoothness compatible with experimental shapers' resolution (e.g., acoustic-optics modulator (AOM), spatial-light modulator (SLM)), we incorporate a total variation denoising (TVD)[64] regularization term $L_{TVD} = \sum |x_{i+1} - x_i|$, with a weighting factor empirically set to $\alpha = 0.5$. The optimization proceeds via iterative backpropagation: in each step, the gradients of the total loss with respect to the input profile are computed through the differentiable surrogate, and the optical waveform is updated using the Adam optimizer until convergence to the global optimum.

**Gradient-based sensitivity map.**

We performed a sensitivity analysis using gradient-based saliency maps. Mathematically, the sensitivity metric $S(t)$ is defined as the absolute magnitude of the gradient of the predicted emittance ε with respect to the instantaneous laser intensity $I(t)$, $S(t) = |\frac{\partial \varepsilon}{\partial I(t)}|$, which quantifies the local susceptibility of the final beam quality to perturbations in the temporal profile. A high value of $S(t)$ indicates that minute variations in intensity at a specific moment $t$ induce significant shifts in the emittance, thereby identifying the temporal regions (such as the pulse rise-time or specific modulation features) that the network prioritizes during optimization.

## Acknowledgment
The author thanks the support from UCLA and SLAC National Accelerator Laboratory, the U.S. Department of Energy (DOE), the Office of Science, Office of Basic Energy Sciences.

## Funding
The U.S. Department of Energy (DOE), the Office of Science, Office of Basic Energy Sciences, under Contract No. DE-AC02-76SF00515, No. DE-SC0022559, No. DE-FOA-0002859, the National Science Foundation under Contract No. NSF2431903, AFOSR FA9550-23-1-0409 and ONR N00014-24-1-2038.

## Declarations

The authors declare no competing interests.

## Contributions

H.Z., R.L., J.H., set up the UV laser shaping experiment and diagnostics. H.Z. designed the computational algorithm, conducted the beamline simulations, and processed the data. H.Z., N.N., A.L.E., J.Q., S.C., performed the data analysis. J.R., T. O. R., D.W., A.M., S.C., provided resources. All authors contributed to interpreting the results and revising the paper.

## Data availability

The relevant code has been publicly accessible via *DOI: 10.5281/zenodo.18943253*.

## Supplementary Information

- **Performance Comparison and Analysis of Surrogate Models**
- **Laser shaping setup**